\documentclass[review]{elsarticle}

\pdfoutput=1

\usepackage{hyperref}
\usepackage{mathtools}
\usepackage{graphicx}
\usepackage[table,xcdraw]{xcolor}
\usepackage{subcaption}
\usepackage{CJKutf8}
\usepackage[bottom]{footmisc}
\usepackage{siunitx}

\usepackage{textcomp}
\usepackage{textgreek}

\renewcommand{\vec}[1]{\mathbf{#1}}

\journal{The 10th International Conference on Management of Emergent Digital EcoSystems (MEDES'18)}

\begin{document}

\begin{frontmatter}

\title{Causal relationship between eWOM topics and profit of rural tourism at Japanese Roadside Stations “MICHINOEKI”}

\author[gidai]{Elisa Claire Alem\'an Carre\'on
\corref{mycorrespondingauthor}}
\ead{s153400@stn.nagaokaut.ac.jp}

\author[gidai]{Tetsuro Ito}
\ead{s151009@stn.nagaokaut.ac.jp}

\author[gidai]{Hirofumi Nonaka}
\ead{nonaka@kjs.nagaokaut.ac.jp}

\author[miyazaki]{Minoru Kumano}
\ead{kumano@cc.miyazaki-u.ac.jp}

\author[nagasaki]{Toru Hiraoka}
\ead{hiraoka@sun.ac.jp}

\author[okayama]{Masaharu Hirota}
\ead{hirota@mis.ous.ac.jp}

\address[gidai]{Nagaoka University of Technology, Nagaoka, Japan}
\address[miyazaki]{University of Miyazaki, Miyazaki, Japan}
\address[nagasaki]{University of Nagasaki, Nagasaki, Japan}
\address[okayama]{Okayama University of Science, Okayama, Japan}

\cortext[mycorrespondingauthor]{Corresponding author}

\begin{abstract}

Affected by urbanization, centralization and the decrease of overall population, Japan has been making efforts to revitalize the rural areas across the country. One particular effort is to increase tourism to these rural areas via regional branding, using local farm products as tourist attractions across Japan. Particularly, a program subsidized by the government called \textit{Michinoeki}, which stands for ‘roadside station’, was created 20 years ago and it strives to provide a safe and comfortable space for cultural interaction between road travelers and the local community, as well as offering refreshment, and relevant information to travelers. However, despite its importance in the revitalization of the Japanese economy, studies with newer technologies and methodologies are lacking. Using sales data from establishments in the Kyushu area of Japan, we used Support Vector to classify content from Twitter into relevant topics and studied their causal relationship to the sales for each establishment using LiNGAM, a linear non-Gaussian acyclic model built for causal structure analysis, to perform an improved market analysis considering more than just correlation. Under the hypotheses stated by the LiNGAM model, we discovered a positive causal relationship between the number of tweets mentioning those establishments, specially mentioning deserts, a need for better access and traffic options, and a potentially untapped customer base in motorcycle biker groups. 

\end{abstract}


\end{frontmatter}

\section{Introduction}\label{intro}

Within the last decades, Japan’s economic growth has been greatly influenced by inbound tourism, creating new potential customers to local businesses across the country and touristic attractions. However, both natural disasters such as earthquakes, as well as the diminishing population in rural areas and its concentration in urban areas have led peripheral areas across the country to attempt regional revitalization via tourism \cite[][]{jones2009}. One of these regional revitalization projects, subsidized by the government after a successful social experiment in the early 1990’s, was created as a way to establish strong links between road users and local communities. This was called ‘Michinoeki’, which stands for ‘Roadside Station’ in Japanese \cite[][]{yokota2006-a}. 

Michinoeki strives to act as a safe and comfortable space in which road travelers can refresh themselves (parking and restrooms); interact with local community and culture, tourist attractions and recreational activities and facilities; as well as provide travelers with relevant information such as maps, emergency care, et cetera. There is currently a network of \num[group-separator={,}]{1145} facilities of Michinoeki across different areas of Japan \cite[][]{michinoeki}. It is important then, in order to revitalize these areas that these facilities match the needs of travelers and tourists. 

\begin{figure}[htp]
\centering
\includegraphics[width=20em]{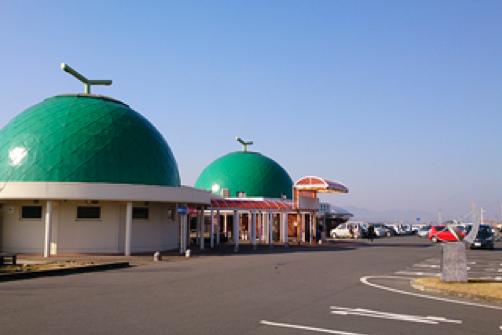}
\caption{Michinoeki Shichijo Melon Dome, in Kukichi city, Kumamoto prefecture, Japan \protect\footnotemark}
\label{fig:melon}
\end{figure}

One notorious example, based in the Kumamoto prefecture in the Kyushu area is ‘Michinoeki Shichijo Melon Dome’, pictured in Figure \ref{fig:melon}. The local product is, of course, melon, and they have used it to make melon taste ‘Melon Pan’ bread (which is named after its shape and not the taste), ice cream and also sold on its own. The place also offers other local souvenirs and products, aside from melon-based ones.

\footnotetext{Shichijo-machi Special Product Center Ltd., Michinoeki Shichijo Melon Dome building, retrieved from \href {http://www.melondome.co.jp/stores_guide/img_stores/melondome_2.jpg}{\path{http://www.melondome.co.jp/stores_guide/img_stores/melondome_2.jpg}}}

Studies pertaining to Michinoeki in Japan are scarce, but within the literary review for this study we found that previous studies, such as one involving the classification of Michinoeki in Hokkaido \cite[][]{ogawa2001}, are mostly based on small surveys. There are additionally studies focusing on the implementation of the idea of Michinoeki in different parts of Asia, such as one in Korea \cite[][]{lee2016}, and another in Vietnam and China \cite[][]{yokota2006-b}. In addition to studies on Michinoeki, there has been focus on the use of regional branding for the revitalization of rural areas by using the local brand farm products as a tourist attraction across Japan \cite[][]{jones2009,ohe2013,ohe2008-a,ohe2008-b}. 

However, in recent years, electronic word-of-mouth (eWOM) has become an important resource for analysis of marketing research and has increasingly been approached by researchers for many products and services \cite[][]{depelsmacker2018,chevalier2006,liu2006}. In addition to this new source of information, Machine Learning approaches, extracting vital information via text mining, among other methodologies of the information age are widely available and increasing in use as well \cite[e.g.][]{he2013,nonaka2012,oconnor2010,Aleman2018ICAROB,Horino2017IEEM,bollen2011,Aleman2017ISIS,nonaka2014icaicta,nonaka2014itmc,nonaka2013,nonaka2010,sakao2009}. These methodologies, when applied to larger databases, provide more trustworthy results than those of statistically small questionnaire samples which can be influenced by the inflexibility of previously posed questions for the customer base. Furthermore, many studies regardless of topic study correlation between the sampled values, leaving to consideration if there is also causation involved. Recently, a Non-Gaussian methodology was presented by Shimizu in 2014, called LiNGAM \cite[][]{shimizu2014}, which allows for a clearer view at the causal relationship between our data.

Because of the importance of the subject of rural tourism to the revitalization of the economy across Japan, and the lack in use of better methodology in this field, we propose to use an Entropy-based Support Vector Machine Learning approach to classify and recognize different topics in eWOM related to Michinoeki extracted from Twitter, and then study the causal relationship between the amount of mentions in those topics and the sales of Michinoeki establishments in the Kyushu area of Japan.

\section{Methodology}\label{methodology}

\subsection{Word Segmentation}\label{segment}

For an analysis to be made possible for each word, we segmented the collected Japanese texts without spaces into words using a Japanese morphological analyzer tool called MeCab \cite[][]{kudo2004}. After segmenting the words, we extracted only self-sufficient words.

\subsection{Entropy Based Keyword Extraction}\label{entropy}

Feature selection of our method is based on the Shannon’s entropy (hereinafter referred as entropy) value \cite[][]{shannon1948} of each word. According to information theory, entropy is the expected value of the information content in a signal.

\begin{figure}[h]
    \centering
    \begin{subfigure}[b]{0.4\linewidth}
        \includegraphics[width=\linewidth]{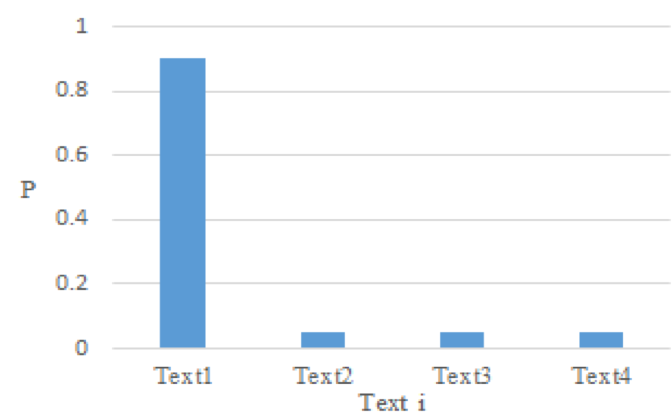}
        \caption{Entropy close to zero.}
    \end{subfigure}
    \begin{subfigure}[b]{0.4\linewidth}
        \includegraphics[width=\linewidth]{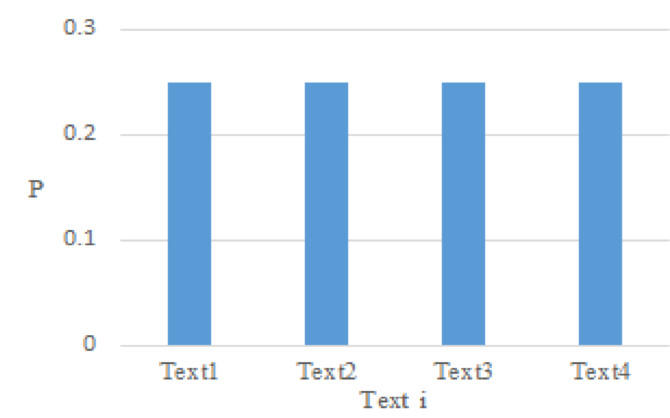}
        \caption{High entropy.}
    \end{subfigure}
\caption{Probabilities of a word \(j\) being contained in a document \(i\).}
\label{fig:entropygraphs}
\end{figure}

Applying this knowledge to the study of words allows us to observe the probability distribution of any given word inside the corpus. For example, a word that keeps reappearing in many different documents will have a high entropy, while a word that only was used in a single text and not in any other documents in the corpus will bear an entropy of zero. This concept is shown in Figure \ref{fig:entropygraphs}.

Having previously tagged a sample of texts positive and negative by pertinence to each category, if a word has higher entropy in positive documents than in negative documents by a factor of alpha greater than 1 (\(\alpha > 1\)), then it means its probability distribution is more spread in positive texts, meaning that it is commonly used in positive tagged documents compared to negative ones.

To calculate the entropy in a set of documents, for each word \(j\) that appears in each document \(i\), we counted the number of times a word appears in positive comments as \(N_{ijP}\), and the number of times a word appears in negative comments as \(N_{ijN}\). Then, as shown in the formulas below, we calculated the probability of each word appearing in each document shown below as \(P_{ijP}\) (\ref{eq:PijP}) and \(P_{ijN}\) (\ref{eq:PijN}).

\begin{equation}\label{eq:PijP}
P_{ijP} = \frac{N_{ijP}}{\sum_{i=1}^M N_{ijP}}
\end{equation}

\begin{equation}\label{eq:PijN}
P_{ijN} = \frac{N_{ijN}}{\sum_{i=1}^M N_{ijN}}
\end{equation}

We then substitute these values in the formula that defines Shannon’s Entropy. We calculated the entropy for each word \(j\) in relation to positive documents as \(H_{Pj}\) (\ref{eq:Hpj}), and the entropy for each word \(j\) in relation to negative texts as \(H_{Nj}\) (\ref{eq:Hnj}). That is, all instances of the summation when the probabilities \(P_{ijP}\) or \(P_{ijN}\) are zero and the logarithm of these becomes undefined are substituted as zero into (\ref{eq:Hpj}) and (\ref{eq:Hnj}).

\begin{equation}\label{eq:Hpj}
H_{Pj} = - \sum_{i=1}^M [P_{ijP}\log_2 P_{ijP}]
\end{equation}

\begin{equation}\label{eq:Hnj}
H_{Nj} = - \sum_{i=1}^M [P_{ijN}\log_2 P_{ijN}]
\end{equation}

After calculating the positive and negative entropies for each word, we measured their proportion using the mutually independent coefficients \(\alpha\) for positive keywords and \(\alpha'\) for negative keywords, for which we applied several values experimentally. A positive keyword is determined when (\ref{eq:entropy_pos}) is true.

\begin{equation}\label{eq:entropy_pos}
H_{Pj} > \alpha H_{Nj}
\end{equation}

\subsection{Topic Classification Using SVM}\label{topics}

In machine learning, Support Vector Machines are supervised learning models commonly used for statistical classification or regression \cite[][]{cortes1995}. We implemented this theory in Python using the Support Vector Classifier (SVC) included in the library \textit{scikit}-learn. We also used the mathematics library \textit{numpy}. To evaluate each of our trained machines, we used the K-fold Cross Validation method, which has been proven to provide good results. To evaluate each of our trained machines, we calculated then the Precision, Recall and \(F_1\)-Score values for our predictions.

\subsection{Causal Relationship Analysis with LiNGAM}

Most previous studies used a linear correlation model to analyze tourism data \cite[e.g.][]{deng2007,koberl2016}, and other traditional methods like the multiple variable regression analysis \cite[][]{thomas2017}, while others have used improved regression models such as the quantile regression to attempt to overcome deficiencies of simple linear correlation models \cite[][]{brida2017}. Correlation suggests an association between two variables. On the other hand, causality shows that one variable directly effects a change in the other. For analysis of tourism industry, it is important to detect cause of sales. Causal structure models have been studied for some time, and there are examples of tourism analysis analyzing predefined causal structures with SEM (Structural Equation Models) \cite[][]{pappas2017}. LiNGAM is currently a widely used model to discover causal structures of continuous-valued data without the need of defining them, under the assumptions that “the data generating process is linear”, “there are no unobserved confounders”, and “disturbance variables have non-Gaussian distributions of non-zero variances”.

The LiNGAM model is as follows:

\begin{equation}\label{eq:lingam_long}
x_i = \sum_{k(j)<k(i)} b_{ij}x_{j} + e
\end{equation}

Where \(k\) represents the causal order of each variable, and in which we are only considering causal orders in the desired direction. This model can also be expressed as:

\begin{equation}\label{eq:lingam1}
\vec{x} = \vec{B}\vec{x} + \vec{e}
\end{equation}

Where the matrix \(\vec{x}\) is comprised of all the measured variables, including in our case the number of tweets and the sales profits; \(e_i\) are continuous latent variables that are exogenous contained in the matrix \(\vec{e}\), and \(b_{ij}\) are the connection strengths from \(x_j\) to \(x_i\). If \(b_{ij}\) is not equal to 0, \(j\) is cause of \(i\). Contrary to this, if \(b_{ij}\) is equal to 0, there is no causality between \(i\) and \(j\), since \(x_i\) would only be comprised of exogenous values (i.e. \(x_i = e_i\)). The matrix \(\vec{B}\) contains all the strength constants and therefore, we must identify it to detect causal structure. Formulation (\ref{eq:lingam2}) can be modified,

\begin{equation}\label{eq:lingam2}
\vec{x} = (\vec{I} - \vec{B})^{-1} \vec{e}
\end{equation}

In order to calculate the matrix \(\vec{B}\), \(\vec{x}\) is expressed in the form of the Independent Component Analysis, or ICA \cite[][]{jutten1991,hyvarinen2001} as follows:

\begin{equation}\label{eq:lingam3}
\vec{x} = \vec{A}_{ICA}\vec{s}
\end{equation}

Where the ICA matrix \(\vec{A}_{ICA}\) collects the coefficients \(a_{ij}\), and \(\vec{s}\) collects the independent components \(s_j\) respectively. However, the output ICA matrix might result in different permutations at the time of calculation. 
Based on the ICA matrix \(\vec{A}_{ICA}\), in order to identify a mixing matrix \(\vec{A}\) such that

\begin{equation}\label{eq:lingam4}
\vec{A} = (\vec{I} - \vec{B})^{-1}
\end{equation}

The matrix \(\vec{A}_{ICA}\) is expressed as

\begin{equation}\label{eq:lingam5}
\vec{A}_{ICA} = (\vec{I} - \vec{B})^{-1} \vec{P}\vec{D} = \vec{A}\vec{P}\vec{D}
\end{equation}

Where \(\vec{P}\) is an unknown permutation matrix and \(\vec{D}\) is an unknown diagonal matrix with no zeros on the diagonal. 
The separating matrix \(\vec{W}\) is defined as

\begin{equation}\label{eq:lingam5}
\vec{W} = \vec{A}^{-1} = \vec{I} - \vec{B}
\end{equation}

Following this, the separating matrix \(\vec{W}\) is estimated up to the permutation \(\vec{P}\), and scaling and sign \(\vec{D}\) of the rows.

\begin{equation}\label{eq:lingam6}
\vec{W}_{ICA} = \vec{P}\vec{D}\vec{W} = \vec{P}\vec{D} \vec{A}^{-1}
\end{equation}

However, in LiNGAM, the correct permutation matrix \(\vec{P}\) can be found \cite[][]{shimizu2006}: the correct \(\vec{P}\) is the only one that contains no zeros in the diagonal of \(\vec{D}\vec{W}\), since \(\vec{B}\) should be a matrix that can be permuted to become lower triangular with all zeros on the diagonal and \(\vec{W} = \vec{I} - \vec{B}\). Furthermore, the correct scaling and signs of the independent components can be determined by using the unity on the diagonal of \(\vec{W} = \vec{I} - \vec{B}\). To obtain \(\vec{W}\) it is only necessary to divide the rows of \(\vec{D}\vec{W}\) by its corresponding diagonal elements. Finally, the connection strength matrix \(\vec{B} = \vec{I} - \vec{W}\) may be computed.

\section{Experiment Results}\label{experiments}

\subsection{Multi-label Topic Classification}\label{exp_topics}

Before the analysis using Support Vector Machines, we defined eight topics by manually. In the sample of tweets related to Michinoeki, we noticed most of the tweets could be classified in these eight topics, with multiple labels in some cases, which are shown in Table 1. We made the training data by sampling 1000 posts and classifying them manually into eight topics. We then calculated the entropy value for each word in each category. With an alpha value of 2, (\(\alpha=2\)) we assigned a word as a keyword for a category only if the entropy value for that category was more than twice the entropy values of all other categories for that word. An exception was made for Topic 5: Check-in, for which the keyword extraction process was done heuristically, choosing keywords such as “I’m at”, indicating only that the user had gone to that particular Michinoeki, which had sufficiently good results.

The topic content and example keywords for each topic obtained from cross comparison of their Entropy values across categories are shown in Table \ref{tab:topics}. 

\begin{table}[htp]
\centering
\caption{Topic Classification Content.}
\label{tab:topics}
\begin{tabular}{|c|l|m{16em}|}
\hline
\rowcolor[HTML]{C0C0C0} 
Topic ID & \multicolumn{1}{c|}{\cellcolor[HTML]{C0C0C0}Topic content} & Example keywords \\ \hline
Topic 1 & Products and Services & shop; set meal; popular; chocolate; ice cream; soft serve; café; \\ \hline
Topic 2 & Special Events & exhibition; event; illumination lighting ceremony; Christmas tree \\ \hline
Topic 3 & Promotional & N/A(Determined by related service company twitter accounts list) \\ \hline
Topic 4 & Traffic and Weather & national highway; attention; road traffic information \\ \hline
Topic 5 & Check-in & I’m at; in; Location \\ \hline
Topic 6 & Positive Reviews & delicious; Instagram-able; cute; cheap; ate; happy; (*{\textasciiacute}\textsuperscript{\textomega}\textasciigrave*) \\ \hline
Topic 7 & Motorcycles & refueling; bike; meeting place; yaeya sticker \\ \hline
Topic 8 & Unrelated and Others & JR; train; wagon; rotary; west exit; ride \\ \hline
\end{tabular}
\end{table}

Because all of the tweets related to Topic 3: Promotional are from official accounts, we categorized these automatically using the usernames without the need to train an SVM or extract entropy-based keywords. Regarding the Topic 9: Unrelated and Others, because of the way we extracted the tweets related to ‘Michinoeki’ which translates to ‘Roadside Stations’, there were cases where actual railroad stations were mentioned instead. We filtered these cases using entropy based SVM as well. Topic 6 describes positive reviews, which we had found differ from other topics; however, little to no negative reviews were found and thus not enough data was available to train a classifier for negative reviews. 

To evaluate each of our trained machines, we calculated the \(F_1\)-Score for each using a K-fold Cross Validation methodology, calculating the \(F_1\)-Score from the values of Precision and Recall calculations. The \(F_1\)-Score values calculated for each SVM are shown in Table \ref{tab:scores}.

\begin{table}[htp]
\centering
\caption{Results of the \(F_1\)-Score in each topic.}
\label{tab:scores}
\begin{tabular}{|c|l|l|}
\hline
\rowcolor[HTML]{C0C0C0} 
Topic ID & \multicolumn{1}{c|}{\cellcolor[HTML]{C0C0C0}Topic content} & \(F_1\)-Score \\ \hline
Topic 1 & Products and Services & 0.85 \\ \hline
Topic 2 & Special Events & 0.81 \\ \hline
Topic 4 & Traffic and Weather & 0.88 \\ \hline
Topic 5 & Check-in & 0.78 \\ \hline
Topic 6 & Positive Reviews & 0.79 \\ \hline
Topic 7 & Motorcycles & 0.87 \\ \hline
Topic 8 & Unrelated and Others & 0.78 \\ \hline
\end{tabular}
\end{table}

\begin{figure}[htp]
\centering
\includegraphics[width=25em]{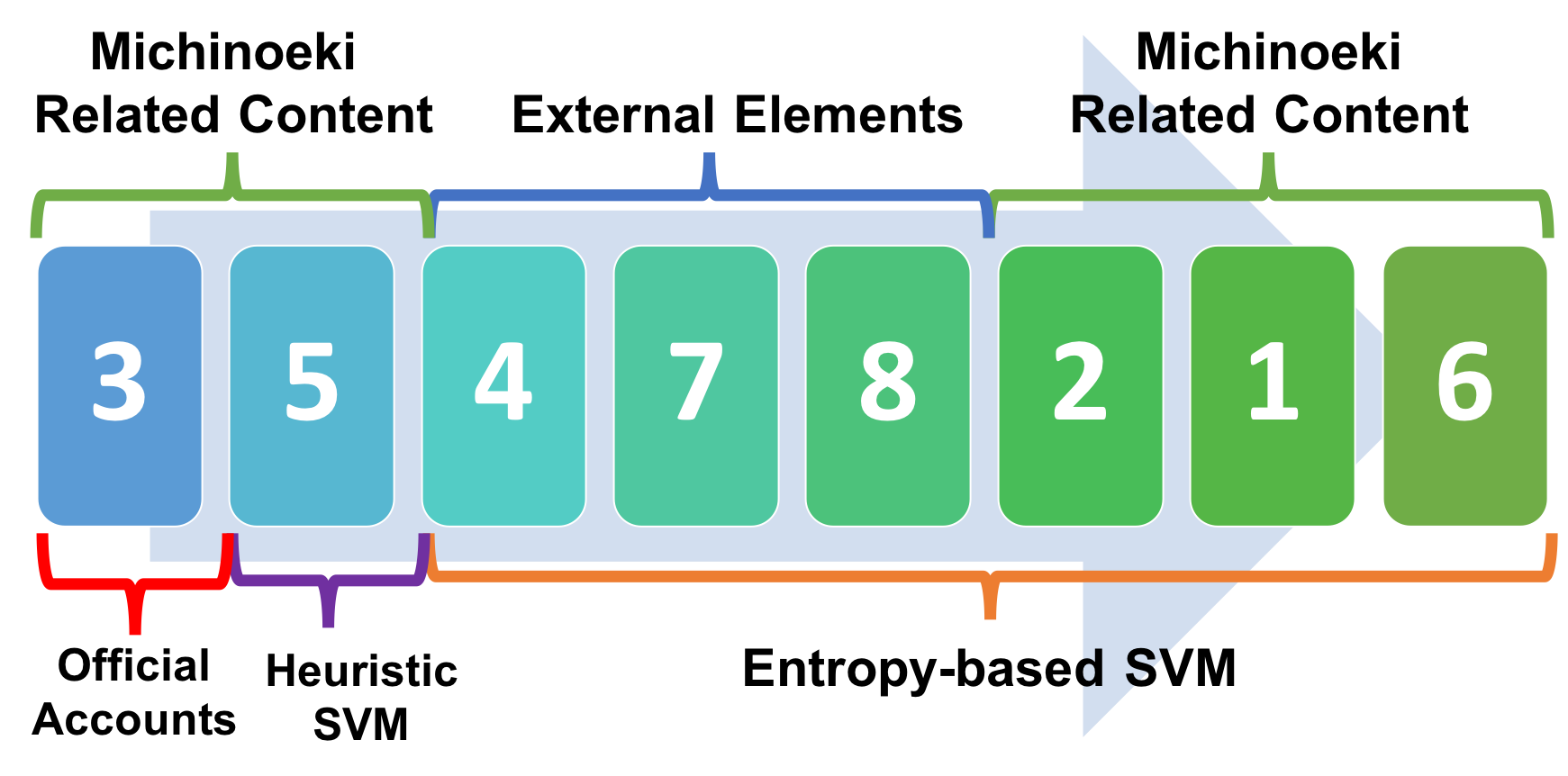}
\caption{Hierarchical topic classification method.}
\label{fig:hierarchical}
\end{figure}

By searching by the names of Michinoeki establishments plus the keyword “michinoeki” in Japanese, we collected a total of \num[group-separator={,}]{111142} tweets related to Michinoeki, of which \num[group-separator={,}]{9264} were related to 94 different Michinoeki establishments across the Kyushu area of Japan for which we had sales data. We then automatically classified posts into one of the eight topics shown on above by using Support Vector Machines in a hierarchical manner. We first classified the tweets belonging to official accounts automatically to Topic 3: Promotional; then classified tweets from Topic 5: Check-In for their particular structure; followed by classification of Topics 4, 7 and 8, which are mostly about external factors and topics unrelated directly with Michinoeki content; finally proceeding to determine positive pertinence to each remaining topic by their respective SVM in order of their \(F_1\)-Score. This heuristic hierarchical binary classification method is shown in Figure \ref{fig:hierarchical}.

\subsection{LiNGAM Analysis Results}\label{res_lingam}

To evaluate the causal relationship between the number of tweets for each of the topics related to Michinoeki as \(x_i\) and the sales of Michinoeki establishments in the Kyushu area of Japan as y, we applied the LiNGAM causal structure analysis to the classified data from the \num[group-separator={,}]{9264} tweets that were matched to those establishments. Results are shown in Table \ref{tab:lingam}.

\begin{table}[htp]
\centering
\caption{Results of the LiNGAM causality analysis.}
\label{tab:lingam}
\begin{tabular}{|l|l|l|l|}
\hline
\rowcolor[HTML]{C0C0C0} 
\multicolumn{1}{|c|}{\cellcolor[HTML]{C0C0C0}Topic ID} & \multicolumn{1}{c|}{\cellcolor[HTML]{C0C0C0}Topic content} & \multicolumn{1}{c|}{\cellcolor[HTML]{C0C0C0}\begin{tabular}[c]{@{}c@{}}Connection \\ Strength\end{tabular}} & \multicolumn{1}{c|}{\cellcolor[HTML]{C0C0C0}\begin{tabular}[c]{@{}c@{}}Causality\\ Direction\end{tabular}} \\ \hline
Topic 1 & Products and Services & 11095.95 & \(x_1 \rightarrow y\) \\ \hline
Topic 2 & Special Events & 515.17 & \(x_2 \rightarrow y\) \\ \hline
Topic 3 & Promotional & 6738.08 & \(x_3 \rightarrow y\) \\ \hline
Topic 4 & Traffic and Weather & -231845.95 & \(x_4 \rightarrow y\) \\ \hline
Topic 5 & Check-in & 1724.36 & \(x_5 \rightarrow y\) \\ \hline
Topic 6 & Positive Reviews & 15387.70 & \(x_6 \rightarrow y\) \\ \hline
Topic 7 & Motorcycles & -7770.03 & \(x_7 \rightarrow y\) \\ \hline
\end{tabular}
\end{table}

\section{Discussion}\label{discussion}

According to the LiNGAM causal structure analysis results, the number of tweets mentioning products and services of Michinoeki establishments, positive reviews of the establishment and their products, as well as promotional tweets published by official accounts and mentions of special events all show a causal relationship with the sales of those establishments. This positive causal relationship is thought to be the influence that twitter mentions have to attract more new customers. As for the tweets from Topic 5: Check-in, there is a clear direct influence, since those users are not only stating that they actually visited and purchased from those establishments, but they are also giving free promotion to the users that follow them. 

Tweets mentioning weather and traffic conditions were observed in our sample to be of mostly expressing inconvenience, which, as is shown by the negative causal constant, has a negative effect on the number of customers that venture to the establishment during those hours, limiting its profit. Tweets in this category also included many complaints about access to the establishments. This shows an opportunity of investment to make clearer signs or routes of access for the affected establishments, as well as marketing campaigns (perhaps using Twitter as well) making their location and access routes more known.

One thing to note is that under inspection of our data, many of the places where there are positive reviews of ice cream and soft serve products are leaders in profitability compared to other establishments. Local ingredients, as well as unique recipes are the main focus of Michinoeki establishments in general, but there could be differences in influence for different kinds of products and specialties. 

Lastly, there is the particular case of the tweets by motorcycle drivers. While Michinoeki establishments strive to be a connection point for road travelers and tourists alike, many bikers use the stations as meeting points only, not contributing to the sales of those establishments. It is necessary in these cases to revise the services provided, such as gasoline stations, for example, so that this untapped customer base can be turned to a positive influence in the future. Another possible strategy could be group campaigns. Most examples in our data of motorcycle drivers were using the place as a meeting point with other motorcycle drivers. Group campaigns or discounts could very well increase their patronage as well as attract more customers in general.

\section{Conclusion and Future Work}\label{conclusion}

We found that the tweets related to Michinoeki could be classified into eight topics with a well performing hierarchical and heuristic approach for multi-class classification using binary SVM classifiers; for which the feature vectors were extracted heuristically in one case, and mathematically in all the other cases by using an entropy-based keyword extraction method. All of the SVM based classifications performed with an \(F_1\)-Score above 0.78, and the highest performing classifier was that of the Topic 4: Traffic and Weather. 

Under the assumptions posed by the LiNGAM model (“the data generating process is linear”, “there are no unobserved confounders”, and “disturbance variables have non-Gaussian distributions of non-zero variances”), we found a causal relationship for all topics and found that most tweets, especially ones praising products, or promoting special events, have a positive influence in sales, with the exception of traffic and weather, and motorcycle travelers, which might be an unexplored market by Michinoeki establishments. However, in future work we will further investigate the assumptions posed by the LiNGAM model in regard to the structure of the data. 

In future work we will study the influence of specific products, their different marketing strategies across different establishments and their relation to their profits in order to discover potential strategies to increase profit in all Michinoeki establishments.

\section*{ACKNOWLEDGMENTS}

This research was supported by Japan Construction Information Center Foundation (JACIC).

\section*{REFERENCES}

\bibliography{michinoeki}

\end{document}